\documentclass[aps,prl,twocolumn,groupedaddress]{revtex4}
\usepackage{graphicx}
\begin{document}

\title{Frenkel Line and Solubility Maximum in Supercritical Fluids}
\author{C. Yang$^{1}$}
\author{V. V. Brazhkin$^{2}$}
\author{M. T. Dove${^1}$}
\author{K. Trachenko${^1}$}
\affiliation{$^1$School of Physics and Astronomy, Queen Mary University of London, Mile End Road, London, E1 4NS, UK}
\affiliation{$^2$Institute for High Pressure Physics, RAS, 142190, Moscow, Russia}

\begin{abstract}
A new dynamic line, the Frenkel line, has recently been proposed to separate the supercritical state into rigid-liquid and non-rigid gas-like fluid. The location of Frenkel line on the phase diagram is unknown for real fluids. Here, we map the Frenkel line for three important systems: CO$_2$, H$_2$O and CH$_4$. This provides an important demarcation on the phase diagram of these systems, the demarcation that separates two distinct physical states with liquid-like and gas-like properties. We find that the Frenkel line can have similar trend as the melting line above the critical pressure. Moreover, we discuss the relationship between unexplained solubility maxima and Frenkel line, and propose that the Frenkel line corresponds to the optimal conditions for solubility.
\end{abstract}


\maketitle

Recently, there has been a dramatical increase of using supercritical fluids in extraction and purification applications, including in food, nuclear waste, petrochemical and pharmaceutical industries \cite{scbook1,scbook2,co2nw1,co2nw2}. Supercritical fluids attract significant attention due to their extremely good dissolving power and ``tunable" properties. The solubility of supercritical fluids depend on density and diffusivity. Supercritical fluids combine the best of both worlds: high density of liquids and large diffusion constants of gases. Moreover, both of those properties can be tuned over a wide range pressure and temperature above the critical point, optimizing their dissolving ability.

Carbon dioxide, water and methane are three most commonly used supercritical fluids. In particular, H$_2$O and CO$_2$,  are both abundant, non flammable and non toxic. They are also ``non-polar'' and ``polar'' solvent, respectively, so they can dissolve ``polar'' and ``non-polar'' solutes, respectively. The critical temperature ($T_{\rm c}$) of CO$_2$ is at 304 K, which is near the room temperature, and the critical pressure ($P_{\rm c}$) is 74 bar, which is also accessible. Additionally, CO$_2$ can be used with co-solvents to modify it into "polar" solvent.

The solubility of variety of solutes have been measured in supercritical CO$_2$ near the $T_{\rm c}$ as a function of pressure \cite{scbook1}. Interestingly, the experiment show intriguing solubility maxima above critical temperature: solubility first substantially increase with pressure, followed by its decrease at higher pressure \cite{solubility1,solubility2,solubility3,solubility4,solubility5,solubility6,solubility7,solubility8}. This effect is not currently understood theoretically. Understanding it would lead to more efficiently use of supercritical fluids. More generally, it is often acknowledged that wider deployment of supercritical fluids and optimizing their use would benefit from a theoretical guidance \cite{scbook1,scbook2}.

Until recently, supercritical state was believed to be physically homogeneous, which means that moving along any path on a pressure and temperature above the critical point does not involve marked or distinct changes. The Frenkel line has recently been proposed, which separates two dynamically distinct states: the gas-like regime where particle only have diffusive motion and the liquid-like regime where particle combine both solid-like quasi-harmonic vibrational motion and gas-like diffusive motion \cite{flpt,fl,fl2}. This transition take place when liquid relaxation time $\tau$ approaches Debye vibration period, $\tau_{\rm D}$. Liquid relaxation time is defined in a usual way as the average time between two consecutive diffusion events (molecular rearrangements between two quasi-equilibrium positions) in the liquid at one point in space \cite{frenkelbook}. When $\tau\approx\tau_{\rm D}$, the system loses the ability to support shear modes at all available frequencies, up to Debye frequency, and retains gas-like diffusive dynamics only. The Frenkel line starts from 0.7--0.8 $T_{\rm c}$ at $P_{\rm c}$ and extends to arbitrarily high pressure and temperature on the phase diagram \cite{flvaf}. There are many ways to locate the Frenkel line on the phase diagram, yet velocity autocorrelation function (VAF) provides a convenient and mathematically meaningful criterion: the disappearance of oscillations and minima of the VAF correspond to pressure and temperature of the Frenkel line \cite{flvaf}.

In this paper, we map the Frenkel line on the phase diagram using Molecular Dynamic (MD) simulation by calculating VAF. We study the location of Frenkel line for CO$_2$, H$_2$O and CH$_4$ on phase diagram, particularly addressing the slop of the Frenkel line in relation to the melting line. We subsequently compare the Frenkel line with the solubility maximum from experiment \cite{solubility1,solubility2,solubility3,solubility5} and discuss why the Frenkel line is related to the solubility maxima.

We use DL\_POLY MD simulation package \cite{dlpoly}, and have simulated 4576 CO$_2$  molecules, 3375 H$_2$O molecules and 3375 CH$_4$ molecules using constant-pressure-temperature ensemble. The intermolecular potential for CO$_2$ is the rigid-body non-polarizable potential based on a quantum chemistry calculation, with the partial charges derived using the Distributed Multipole Analysis method \cite{co2potential}. The intermolecular potential of H$_2$O is TIP4P/2005, which can describe intermolecular force very well \cite{tip4p}. The intermolecular potential of CH$_4$ is taken from Refs \cite{ch4potential}. This potential also shows good accuracy in the supercritical state. We used cut-off as 12 \AA for potential, and the Smoothed Particle Mesh Ewald for long-range forces. We first equilibrate the system during 10 ps, and ensure the equilibration at given ($P,T$) conditions during the subsequent 40 ps. We collect and analyse the result during following 50 ps. In the range of our simulations, the difference between MD and experimental density from the NIST data base \cite{nist}, is less than 5\%. Our pressure range extends to several GPa and includes the pressure used in industrial applications.

\begin{figure}[!]
\begin{center}
{\scalebox{0.325}{\includegraphics{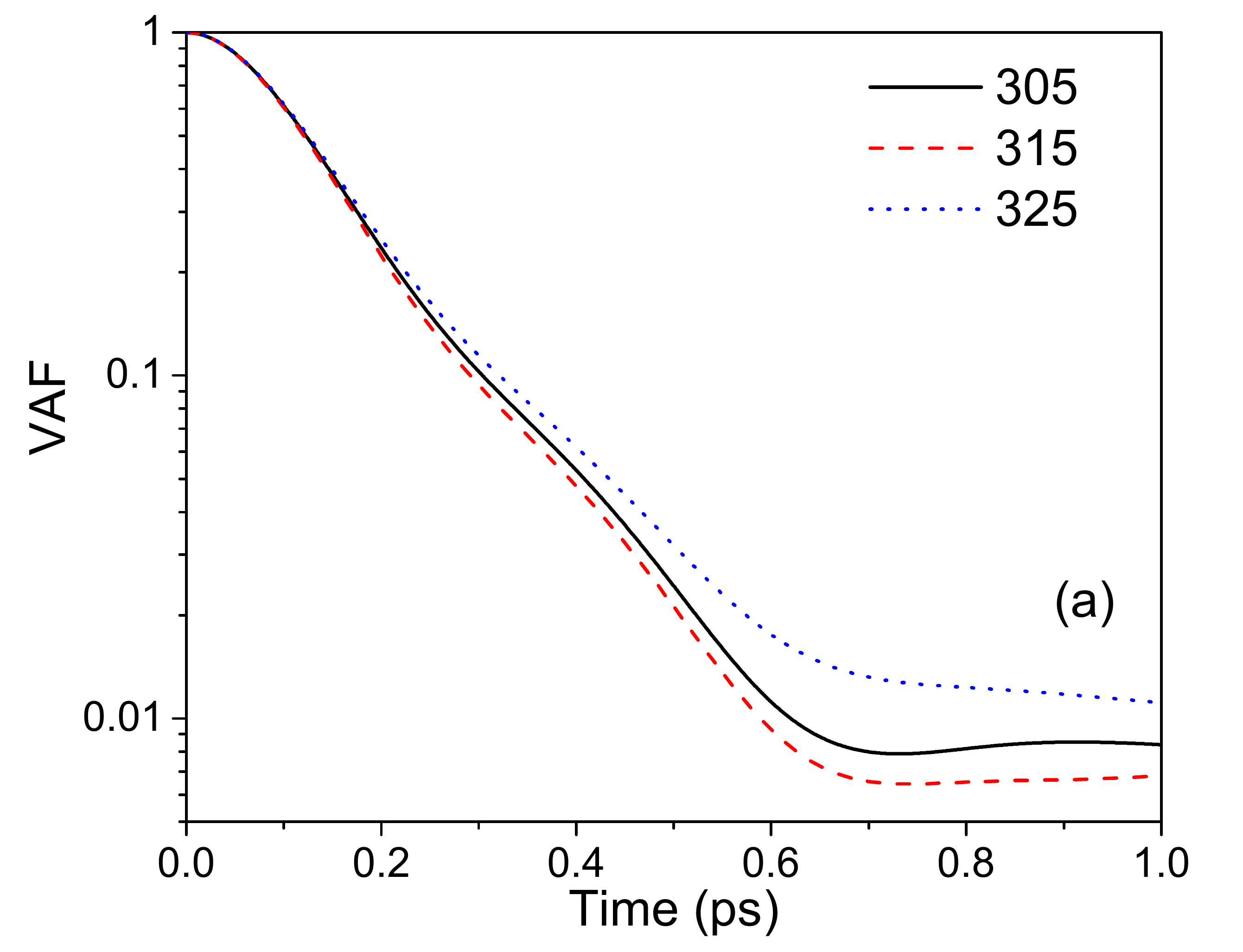}}}
\end{center}
\begin{center}
{\scalebox{0.31}{\includegraphics{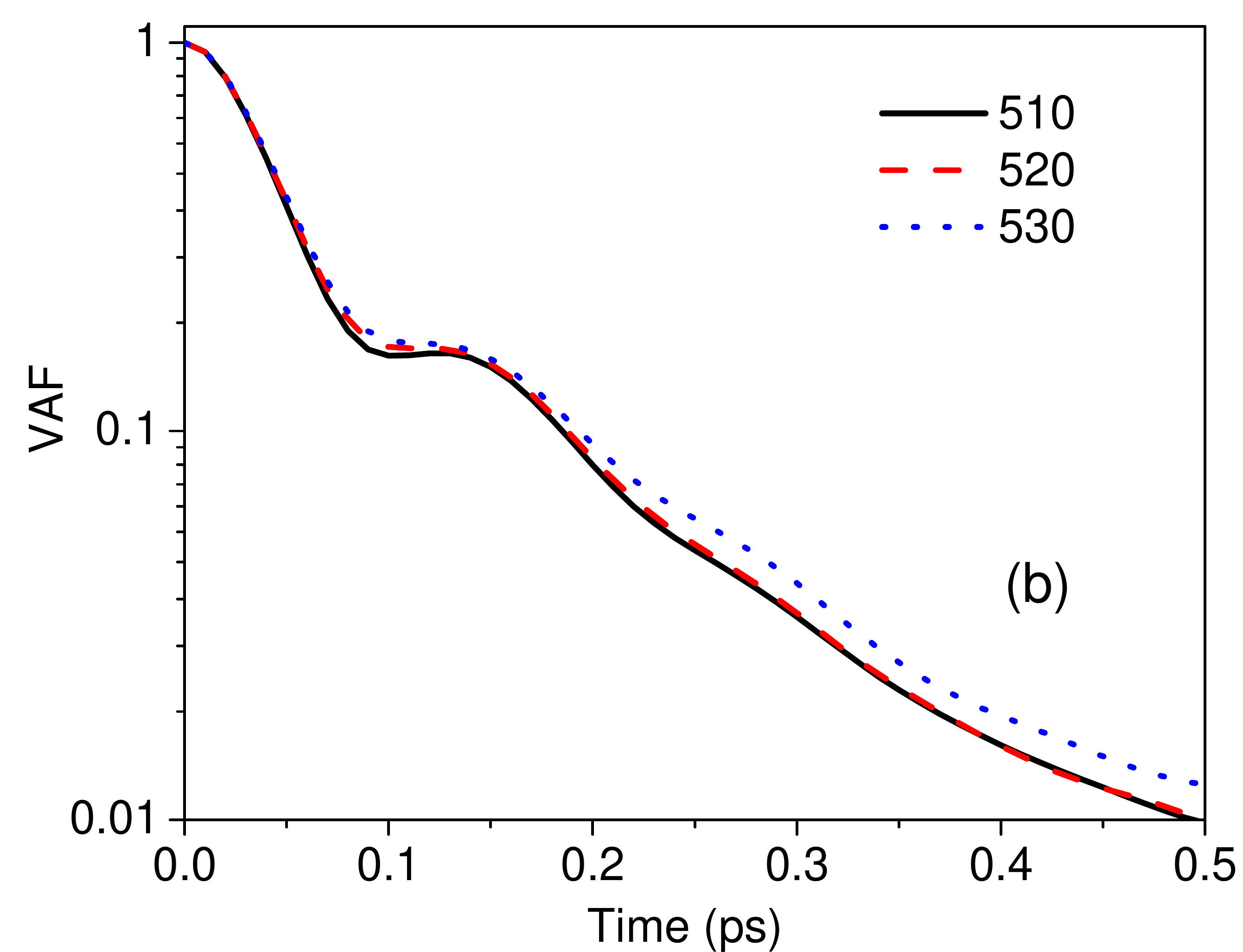}}}
\end{center}
\begin{center}
{\scalebox{0.31}{\includegraphics{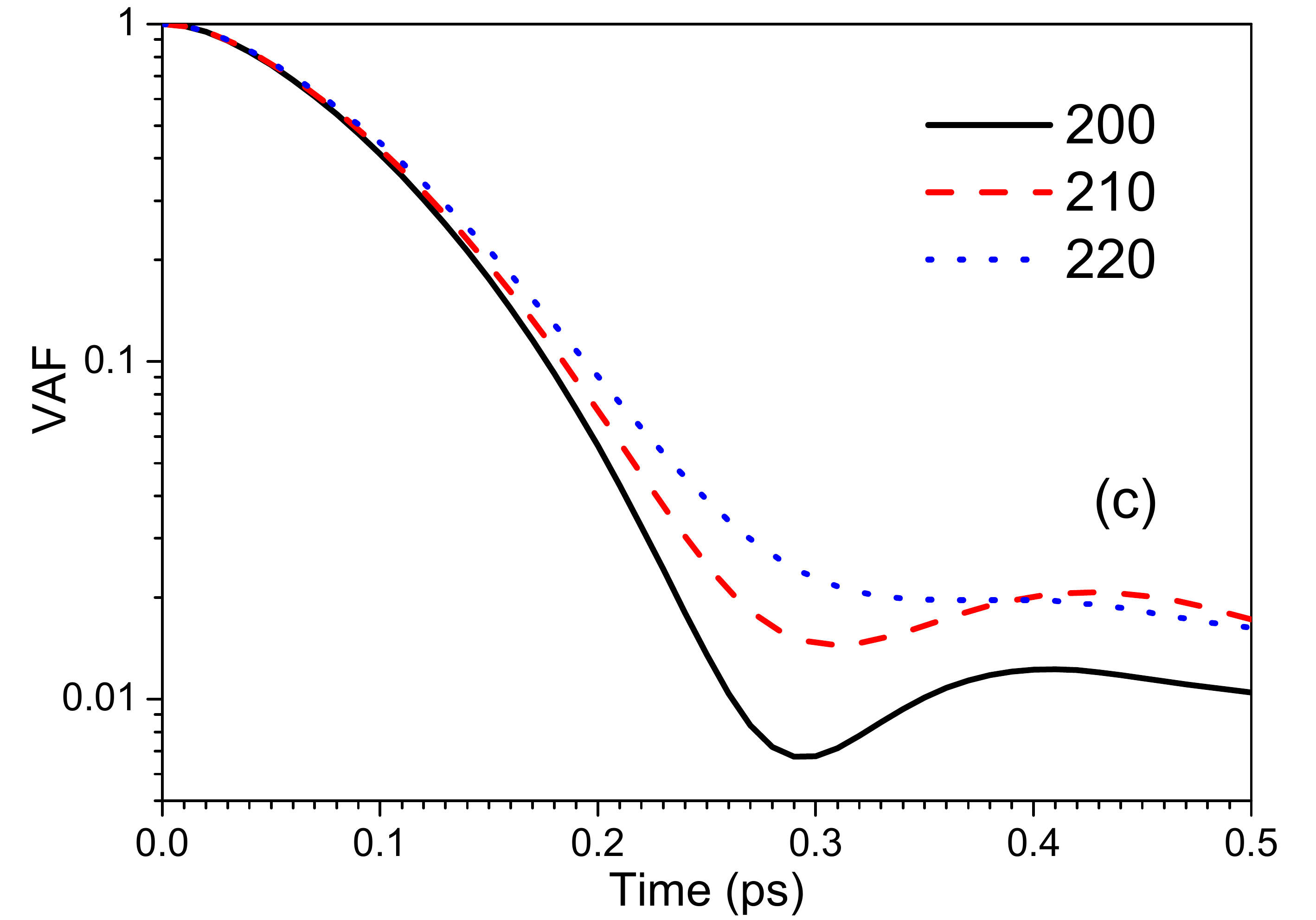}}}
\end{center}
\caption{(Color online)Velocity auto-correlation functions for CO$_2$ (a), H$_2$O (b) and CH$_4$ (C) showing the crossover at the Frenkel line at 900 bar.
}
\label{1}
\end{figure}
It is well known that VAF is a monotonically decaying function in the gas state, whereas it shows damped oscillation in the liquid and solid state. VAF is defined as:
\begin{equation}
Z(t)=<\vec{v}(0)\cdot \vec{v}(t)>
\end{equation}

In the previous work \cite{flvaf}, it was shown that the minimum of VAF would disappear when the system crosses the Frenkel line in the supercritical state. In Fig. 1, we show representative VAFs for CO$_2$, H$_2$O and CH$_4$ at 900 bar. We can clearly see that as the temperature increases, the minimum becomes more shallow and finally disappears, which corresponds to the loss of oscillatory component of molecular motion and gives ($P$,$T$) for the Frenkel line. We note that the longitudinal mode persists above the Frenkel line, albeit starts disappearing with temperature, starting with the shortest wavelengths \cite{natcomm}.

In Fig. 2, we map the Frenkel line for CO$_2$, H$_2$O and CH$_4$ using the VAF criterion (disappearance of the first minimum). For technologically important CO$_2$ and H$_2$O, we show the Frenkel line in both (pressure, temperature) and (density, temperature) coordinates. We also show the melting line \cite{mtco1,mtco2,mtm1,mtm2} on the phase diagram.

We observe that the Frenkel line for all three fluids starts from 0.7$\sim$ 0.8 $T_{\rm c}$ at $P_{\rm c}$, which agrees with our previous result on Lennard-Jones fluids \cite{flvaf}. Notably, the Frenkel line does not need to start from the critical point because fundamentally it is related to critical phenomena, and exists in systems such as the soft-sphere system where the boiling line and the critical point are absent altogether \cite{flvaf}.

\begin{figure*}[!]
\begin{center}
{\scalebox{0.34}{\includegraphics{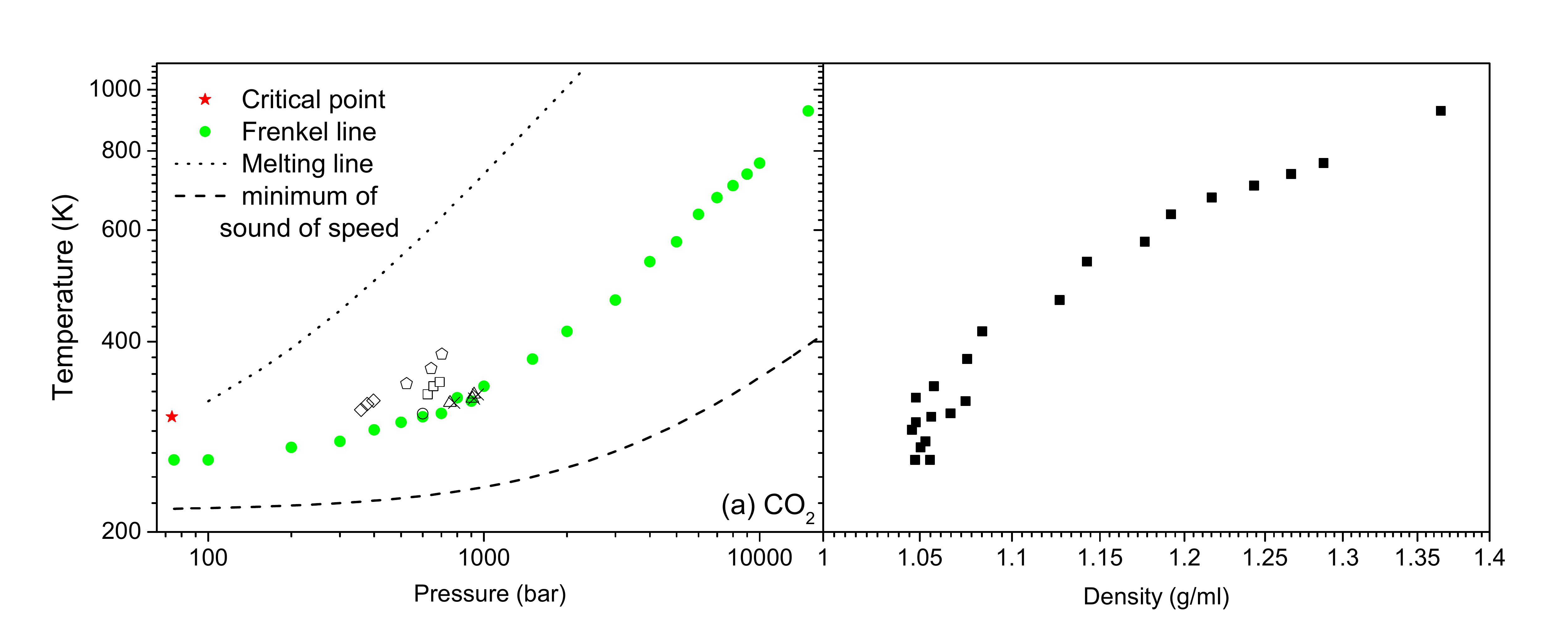}}}
\end{center}
\begin{center}
{\scalebox{0.34}{\includegraphics{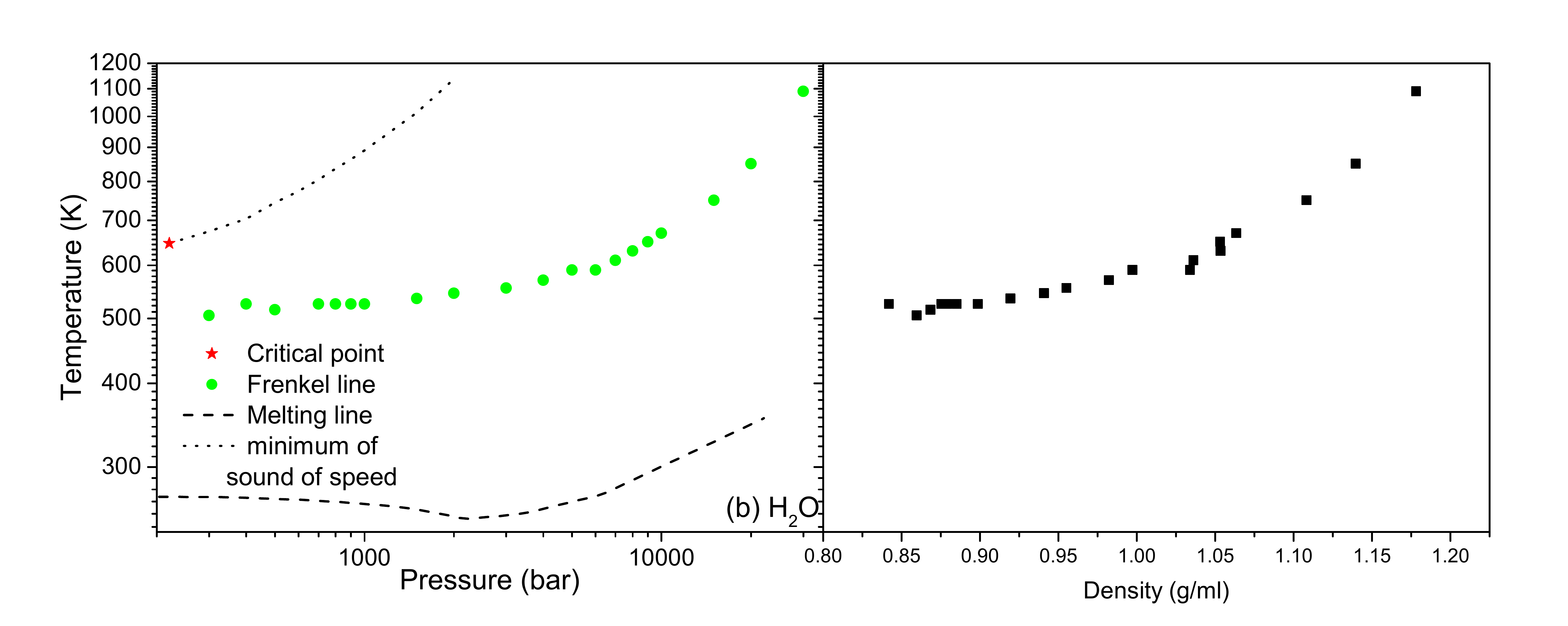}}}
\end{center}
\begin{center}
{\scalebox{0.34}{\includegraphics{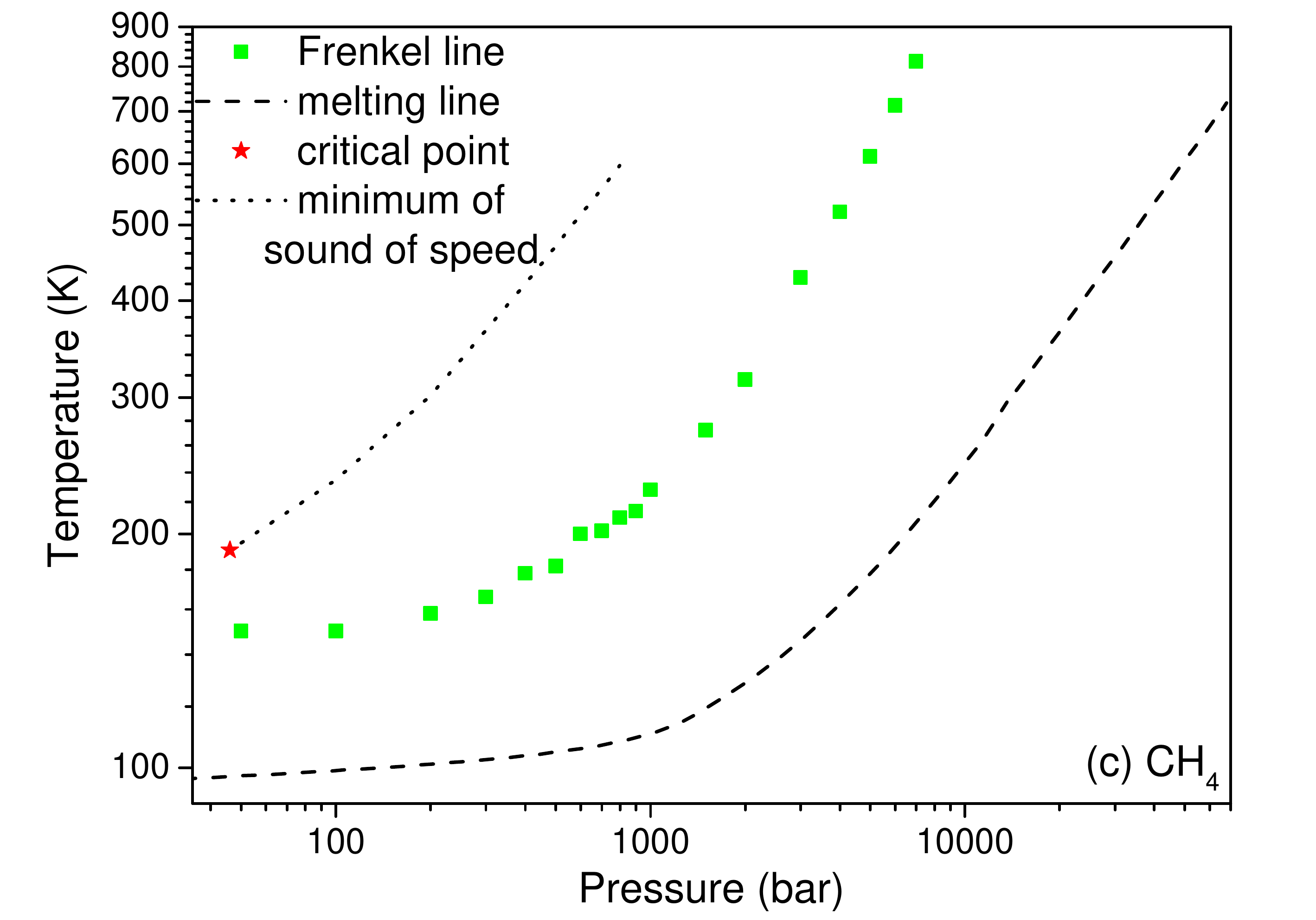}}}
\end{center}
\caption{(Color online)The Frenkel line for CO$_2$, H$_2$O on the pressure-temperature (left) and density-temperature (right) phase diagram. The Frenkel line for CH$_4$ is shown in the pressure-temperature diagram. The solubility maximum of different solutes in supercritical CO$_2$ are shown in graph (a). The open circle are the solubility of $\beta$-carotene \cite{solubility5}; the squares are 1,4-bis-(octadecylamino)-9,10-anthraquinone \cite{solubility1}; triangles are 1,4-bis-(n-alkylamino)-9,10-anthraquinone \cite{solubility2};  the diamonds are biphenyl \cite{solubility3}; the pentagon are adamantane \cite{solubility6}; the cross are 1,4-bis-(hexadecylamino)-9,10-anthraquinone \cite{solubility7}
}
\label{2}
\end{figure*}

We now discuss a relationship between the Frenkel line and the melting line, the relationship that can serve as a useful guide to map the Frenkel line on the phase diagram for any system. As discussed above, the Frenkel line starts slightly below the critical point. At higher pressure, we can predict that the Frenkel line is parallel to the melting line in the log-log plot on the basis of the well-known scaling argument. Indeed, starting from high GPa pressures, the intermolecular interaction is reduced to its repulsive part only, whereas the cohesive attracting part no longer affects interactions (at low pressure, the parallelism between the two lines holds only approximately because the interactions are not well approximated by simple repulsive laws, see below). In a sufficiently wide pressure range, the repulsive part can be well approximated by several empirical interatomic potentials such as the Buckingham-type functions or Lennard-Jones potentials with inverse power-law leading terms at short distances $U \propto \frac{1}{r^n}$ . For the inverse power law, a well-known scaling of pressure and temperature exists: the system properties depend only on the combination of $TP^\gamma$ , where $\gamma$ is uniquely related to n. Consequently, $TP^\gamma$ = const. on all ($P,T$) lines where the dynamics of particles changes qualitatively, as it does on both the melting line and the Frenkel line. This implies that the Frenkel and melting lines are parallel to each other in the double-logarithmic plot, the insight that we have recently used to construct the Frenkel line for molecular hydrogen \cite{flh2}.

Although our simulations were in the practically useful range of pressure and did not extend to high enough pressure to meet the condition above, we observe that the Frenkel line has similar trend as the melting line: for CO$_2$, the slopes of the Frenkel line and the melting line both starts to increase around 1,000 bar. For H$_2$O, both lines are flat below 1,000 bar, but their slopes start increasing at higher pressure. We also observe similar slope increase for methane simultaneously around 1,200 bar.

The speed of sound is one of important properties shows qualitatively changes in the supercritical state. Notably, the speed of sound decreases with temperature below the Frenkel line, as in liquids and solids, but increases with temperature sufficiently above the line, as in gases \cite{flpt,fl,fl2}. We note that the minima of the speed of sound are not absolute in the sense that their positions depend on the path on the diagram (the position of the minimum along isobars, isochors and isotherms can be different). The scaling argument above implies that if the minima of the speed of sound correspond to the qualitative change of particle dynamics, the line of these minima should be approximately parallel to the Frenkel line. In Fig.2, we show pressure and temperature that correspond to the minimum of the speed of sound as deduced from the NIST database\cite{nist}. We observe that the line of speed of sound minimum is approximately parallel to the Frenkel line at high pressure as predicted.

To discuss the relationship between the solubility and the dynamic property of supercritical fluid, we show isothermal solubility maxima of different solutes in CO$_2$ \cite{solubility1,solubility2,solubility3,solubility4,solubility5,solubility6,solubility7}, on the phase diagram (Fig. 2(a)). Importantly, we observe the points of solubility maxima are close to the Frenkel line. The solubility of maxima of several solutes, such as $\beta$-carotene, 1,4-bis-(n-alkylamino)-9,10-anthraquinone and 1,4-bis-(hexadecylamino)-9,10-anthraquinone coincide with the Frenkel line.

We now explain the proximity of the solubility maxima and Frenkle line, as follows. Let us fix a temperature above the critical point to the left of the Frenkel line and increase the pressure (moving horizontally to the right in Fig. 2). Pressure has two competing effects on diffusion. On one hand, it increases density and hence the contact area and cleaning (dissolving) efficiency. On the other hand, the density increase results in decreasing the diffusion constant. Indeed at the Frenkel line, where the molecular dynamics acquires the oscillatory component, molecular rearrangements become markedly less frequent, in contrast to the gas-like dynamics above the line where the oscillatory component of motion is absent. Therefore, at the Frenkel line, the supercritical fluid has maximal density possible at which the diffusion is still in the fast gas-like regime and not in the slow liquid-like regime. The optimum combination of these two properties gives solubility maxima.

The data for solubility maximum for H$_2$O and CH$_4$ are not available. From the Fig. 2 (b) and (c), we see the reason why it is difficult to perform these in experiments. In the case of water, the pressure of Frenkel line is $\sim$9,000 bar at $T_{\rm c}$, which is $\sim$40 times of $P_{\rm c}$ of H$_2$O (220.64 bar). Although, the pressure of CH$_4$ is not too high ($\sim$600 bar) at $T_{\rm c}$, the pressure increase to $\sim$4,000 bar at room temperature. In both cases, their pressure are much higher than $P_{\rm c}$. Compared with them, CO$_2$ is located $\sim$1,000 bar at $T_{\rm c}$, which is relatively more affordable in experiment. We propose the Frenkel line serve as a predictive tool to locate the solubility maxima on the phase diagram. This provides a useful guide for future experiments.

We note that the increase of pressure along the Frenkel line results in several other interesting and potentially important effects such as the increase of fluid density and diffusion constant as well as the appearance of the viscosity minimum \cite{flpt,fl,fl2,flvaf}. In addition, surface tension tends to zero around and above the critical point so that that the problem of wetting is avoided. Accordingly, these conditions may favour other important properties of supercritical fluids: for example, the speed of chemical reactions may have a maximum close to the Frenkel line. In this and other cases, supercritical technology will further benefit from theoretical guidance and receive an impetus for using the supercritical fluids in the hitherto unknown range of 1-10 kbars.

In summary, we mapped the Frenkel line for three important system: CO$_2$, H$_2$O and CH$_4$. This provides an important demarcation on the phase diagram of these systems, the demarcation that separates two distinct physical states with liquid-like and gas-like properties. We proposed that the Frenkel line can serve as a important guide to estimate the location of solubility maxima, so that the cleaning and dissolving abilities of the supercritical fluids are optimized.

K. T. is grateful to EPSRC, C. Yang to CSC. V. V. Brazhkin is grateful to RSF (14-22-00093) for the financial support.

\end{document}